\documentclass[12pt,a4paper]{article}
\usepackage{a4wide}
\usepackage{latexsym}
\usepackage{epsf}
\usepackage{amssymb}

\makeatletter
\@addtoreset{equation}{section}
\makeatother

\begin{document}

\begin{flushright}
\small
IFT-UAM/CSIC-02-35\\
{\bf hep-th/0208158}\\
August $21$st, $2002$
\normalsize
\end{flushright}

\begin{center}


\vspace{2cm}

{\Large {\bf Geometric Construction of Killing Spinors and}}

\vspace{.5cm}

{\Large {\bf Supersymmetry Algebras in Homogeneous Spacetimes}}

\vspace{2cm}

{\bf\large Natxo Alonso-Alberca}${}^{\spadesuit\heartsuit}$
\footnote{E-mail: {\tt natxo@leonidas.imaff.csic.es}},
{\bf\large Ernesto Lozano-Tellechea}${}^{\spadesuit\heartsuit}$
\footnote{E-mail: {\tt Ernesto.Lozano@uam.es}}\\
{\bf\large and Tom{\'a}s Ort\'{\i}n}${}^{\spadesuit\clubsuit}$
\footnote{E-mail: {\tt Tomas.Ortin@cern.ch}}

\vspace{1cm}

${}^{\spadesuit}$\ {\it Instituto de F\'{\i}sica Te{\'o}rica, C-XVI,
Universidad Aut{\'o}noma de Madrid\\
Cantoblanco, E-28049 Madrid, Spain}

\vskip 0.2cm
${}^{\heartsuit}$\ {\it Departamento de F\'{\i}sica Te{\'o}rica, C-XI,
Universidad Aut{\'o}noma de Madrid\\
Cantoblanco, E-28049 Madrid, Spain}

\vskip 0.2cm
${}^{\clubsuit}$\ {\it I.M.A.F.F., C.S.I.C., 
Calle de Serrano 113 bis\\ 
E-28006 Madrid, Spain}
\vspace{.7cm}


{\bf Abstract}

\end{center}

\begin{quotation}

\small

We show how the Killing spinors of some maximally supersymmetric
supergravity solutions whose metrics describe symmetric spacetimes
(including $AdS,AdS\times S$ and H$pp$-waves) can be easily
constructed using purely geometrical and group-theoretical methods.
The calculation of the supersymmetry algebras is extremely simple in
this formalism.

\end{quotation}

\newpage

\pagestyle{plain}


\section*{Introduction}

In theories with local supersymmetry (supergravity and superstring
theories), the maximally supersymmetric solutions are usually
identified as vacua, although vacua with less unbroken supersymmetry
can also be interesting. The vacuum supersymmetry algebra, together
with Wigner's theorem, determine which fields can be defined on it,
their conserved (quantum) numbers, the particle spectrum etc. Thus,
the supersymmetry algebra is a very important piece of information.

The calculation of the supersymmetry algebra of a solution (see,
e.g.~Ref.~\cite{Figueroa-O'Farrill:1999va}), involves the calculation
of its Killing vectors and Killing spinors, and the computation of
bilinears and Lie derivatives of the Killing spinors which can
sometimes be difficult or tedious, since their functional form has no
manifest geometrical meaning.

However, most known maximally supersymmetric solutions have the
spacetime metric of some symmetric space that can be identified with a
coset $G/H$\footnote{The only exception known to us could be the
  near-horizon metric of the 5-dimensional rotating black hole.}.  Our
main result is that, quite generally, the Killing spinor equation in
maximally supersymmetric solutions can be put in the form

\begin{equation}
\label{eq:KSEu}
(d +u^{-1} du)\kappa=0\, , 
\end{equation}

\noindent 
which, written in the form $u^{-1}d(u\kappa)=0$ tells us that the
Killing spinors are given by 

\begin{equation}
\kappa = u^{-1}\kappa_{0}\, ,  
\end{equation}

\noindent
where $\kappa_{0}$ is a constant Killing spinor. $u$ is a coset
representative in the spinorial representation.  Then, the bilinears
$\bar{\kappa}\gamma^{\mu}\kappa$ can be easily decomposed into Killing
vectors and the Lie-Lorentz derivative of the Killing spinors with
respect to the Killing vectors are also easily computed. This
simplifies dramatically the calculation of the supersymmetry algebras
of these maximally supersymmetric solutions.

In Section~\ref{sec-symmetric} we give a extremely sketchy review of
the theory of symmetric spaces needed to prove the above general
result in the examples that will follow.  In Section~\ref{sec-KSSS} we
use the machinery just introduced to give a construction of the metric
of several well-known maximally supersymmetric supergravity solutions
(all of them corresponding to symmetric, but, in general, not
maximally symmetric spacetimes) and to show how the general rule for
the construction of the Killing spinors works in practice. We start
with the simplest non-trivial example: $AdS_{4}$ in $N=1,d=4$ ($AdS$)
supergravity (Section~\ref{sec-AdSKS}). Then we consider the next
non-trivial example: the Robinson-Bertotti solution with geometry
$AdS_{2}\times S^{2}$ (Section~\ref{sec-RB}) which we then generalize
to other known maximally supersymmetric solutions with geometries of
the type $AdS\times S$ (Section~\ref{sec-AdSxS}).  Finally, we
consider in Section~\ref{sec-KSSS} the last kind of known maximally
supersymmetric solutions: the Kowalski-Glikman solutions with
H$pp$-wave geometries.  Section~\ref{sec-conclusions} contains our
conclusions perspectives for future work.


\section{Symmetric Spaces}
\label{sec-symmetric}

Let us consider\footnote{Two Physics-oriented general references are
  \cite{Castellani:et} and \cite{Coquereaux:ne}.} the
$(p+q)$-dimensional Lie group $G$, its $p$-dimensional subgroup $H$
and the $q$-dimensional space of right cosets $G/H=\{gH\}$. The Lie
algebra $\mathfrak{g}$ of $G$ is spanned by the generators $T_{I}$
($I=1,\cdots,p+q$) with Lie algebra

\begin{equation}
[T_{I},T_{J}] = f_{IJ}{}^{K}T_{K}\, .  
\end{equation}

\noindent
The Lie algebra of $H$ is generated by the subalgebra
$\mathfrak{h}\subset \mathfrak{g}$ spanned by the generators $M_{i}$
($i=1,\cdots,p$) with Lie brackets

\begin{equation}
[M_{i},M_{j}] = f_{ij}{}^{k}M_{k}\, .  
\end{equation}

The vector subspace spanned by the remaining generators, denoted by
$P_{a}$, ($a,b=1,\cdots,q$) is denoted by $\mathfrak{k}$ and, as vector
spaces we have $\mathfrak{g}=\mathfrak{k}\oplus\mathfrak{h}$.
Exponentiating the generators of $\mathfrak{k}$ we can construct a
coset representative $u(x)=u(x^{1},\cdots,x^{q})$. We will always
construct the coset representative as a product of generic elements of
the 1-dimensional subgroups generated by the $P_{a}$s:

\begin{equation}
\label{eq:cosetrepresentative}
u(x) =  e^{x^{1}P_{1}}\cdots  e^{x^{q}P_{q}}\, .  
\end{equation}

Under a left transformation $g\in G$ $u$ transforms into another
element of $G$ which only becomes a coset representative
$u(x^{\prime})$ after a right transformation with an element $h\in H$,
which is a function of $g$ and $x$:

\begin{equation}
gu(x) = u(x^{\prime}) h\, .  
\end{equation}

The adjoint representation of $\mathfrak{g}$ has as representation
space $\mathfrak{g}$ and can be defined by its action on its
generators: for any $T\in\mathfrak{g}$

\begin{equation}
\Gamma_{\rm Adj}\, (T) (T_{I})\equiv [T,T_{I}]\, ,
\,\,\,\, \Rightarrow 
\Gamma_{\rm Adj}\, (T_{I})^{K}{}_{J}= f_{IJ}{}^{K}\, .
\end{equation}

Exponentiating the generators of the Lie algebra $\mathfrak{g}$ in the
adjoint representation, we get the adjoint representation of the group
$G$

\begin{equation}
\Gamma_{\rm Adj}\,(g(x))
=\exp{\{x^{I}\Gamma_{\rm Adj}\,(T_{I})\}}  \, .
\end{equation}

\noindent that acts on the Lie algebra generators

\begin{equation}
T^{\prime}_{J} = T_{L}\Gamma_{\rm Adj}\,(g)^{L}{}_{J} \, .
\end{equation}

Actually, in any representation $r$, the adjoint action of $G$ on
$\mathfrak{g}$ is given by

\begin{equation}
\Gamma_{r}(g)\Gamma_{r}(T_{I})\Gamma_{r}(g^{-1}) = 
\Gamma_{r}(T_{J}) \Gamma_{\rm Adj}\,(g)^{J}{}_{I}\, . 
\end{equation}

The {\it Killing metric} $K_{IJ}$ is defined by 

\begin{equation}
K_{IJ} \equiv {\rm Tr} 
[\Gamma_{\rm Adj}\,(T_{I}) \Gamma_{\rm Adj}\,(T_{J}) ]\, ,  
\end{equation}

\noindent
and by construction it is invariant under the adjoint action of $G$,
due to the cyclic property of the trace.

The homogeneous space $G/H$ can be used to construct a symmetric space
if the pair $(\mathfrak{k},\mathfrak{h})$ is a {\it symmetric pair}
satisfying

\begin{equation}
  \begin{array}{rcl}
\left[\mathfrak{h},\mathfrak{h} \right] & \subset & \mathfrak{h}\, ,\\
& & \\
\left[\mathfrak{k},\mathfrak{h} \right] & \subset & \mathfrak{k}\, ,\\
& & \\
\left[\mathfrak{k},\mathfrak{k} \right] & \subset & \mathfrak{h}\, .\\
  \end{array}
\end{equation}

\noindent
The first condition is always satisfied for homogeneous spaces since
$\mathfrak{h}$ is a subalgebra. The second condition says that
$\mathfrak{k}$ is a representation of $H$. The two components of a
symmetric pair are mutually orthogonal with respect to the Killing
metric which is block-diagonal.

The first step is the construction of the left-invariant
Lie-algebra valued {\it Maurer-Cartan 1-form} $V$

\begin{equation}
\label{eq:M-Cdef}
V \equiv - u^{-1}du =  e^{a}P_{a}+\vartheta^{i}M_{i}\, ,  
\end{equation}

\noindent
that we have decomposed in {\it horizontal} $e^{a}$ and {\it vertical}
components $\vartheta^{i}$. By construction, $V$ satisfies the {\it
  Maurer-Cartan equations}

\begin{equation}
dV- V\wedge V=0\, ,
\,\,\,\,
\Rightarrow\,\,
\left\{
  \begin{array}{lcl}
de^{a} -\vartheta^{i}\wedge e^{b} f_{ib}{}^{a} & = & 0\, ,\\
& & \\
d\vartheta^{i} 
-\frac{1}{2}\vartheta^{j}\wedge 
\vartheta^{k} f_{jk}{}^{i}
-\frac{1}{2}e^{a}\wedge e^{b}  
f_{ab}{}^{i} & = & 0\, .\\
  \end{array}
\right.  
\end{equation}

The horizontal components $e^{a}$ provide us with a co-frame for
$G/H$.  Under left multiplication by a constant element $g\in G$
$u(x^{\prime}) = gu(x)h^{-1}$, which implies for the Maurer-Cartan
1-form components

\begin{equation}
\label{eq:transformation1}
\left\{
  \begin{array}{rcl}
e^{a}(x^{\prime}) & = & (h e(x) h^{-1})^{a} 
= \Gamma_{\rm Adj}(h)^{a}{}_{b} e^{b}(x)\, ,\\
& & \\
\vartheta^{i}(x^{\prime}) & = & 
(h \vartheta(x) h^{-1})^{i} +(h^{-1}dh)^{i}\, .\\
  \end{array}
\right.
\end{equation}

The second step to construct a symmetric space is the construction of
the metric. With a symmetric bilinear form $B_{ab}$ in $\mathfrak{k}$
we can construct a Riemannian metric

\begin{equation}
ds^{2} \sim B_{ab}e^{a}\otimes e^{b}\, ,  
\end{equation}

\noindent 
that will be invariant under the left action of $G$ if $B$ is:

\begin{equation}
f_{i(a}{}^{c}B_{b)c}=0\, .  
\end{equation}

\noindent 
This is guaranteed if $B_{mn}=K_{mn}$, the projection on
$\mathfrak{k}$ of the Killing metric, but sometimes this is singular
and another one has to be used . The resulting Riemannian metric
contains $G$ in its isometry group (which could be bigger) and must
admit $p+q$ Killing vector fields $k_{(I)}$. The Killing vectors
$k_{(I)}$ and the so-called {\it $H$-compensator} $W_{I}{}^{i}$ are
defined through the infinitesimal version of the transformation rule
$gu(x)=u(x^{\prime})h$ with

\begin{equation}
  \begin{array}{rcl}
g & = & 1 + \sigma^{I}T_{I}\, ,\\
& & \\
h & = & 1 -\sigma^{I}W_{I}{}^{i}M_{i}\, ,\\
& & \\
x^{\mu\, \prime} & = & x^{\mu} +\sigma^{I}k_{(I)}{}^{\mu}\, .\\
  \end{array}
\end{equation}

\noindent 
Using the above equations into

\begin{equation}
u(x+\delta x)= u(x) + \sigma^{I} k_{(I)}u\, ,
\end{equation}

\noindent 
we get 

\begin{equation}
\label{eq:esa}
T_{I}u = k_{I}u -uW_{I}{}^{i}M_{i}\, .
\end{equation}

\noindent 
Acting with $u^{-1}$ on the left and using the definitions of the
adjoint action and the Maurer-Cartan 1-forms , we get

\begin{equation}
T_{J}\Gamma_{\rm Adj}(u^{-1})^{J}{}_{I} = -k_{(I)}{}^{a}P_{a} 
-(k_{(I)}{}^{\mu}\vartheta^{i}{}_{\mu} +W_{I}{}^{i})M_{i}\, ,   
\end{equation}

\noindent
which, projected on the horizontal and vertical subspaces gives the
following expressions for the tangent space components of the Killing
vector fields and the $H$-compensator

\begin{eqnarray}
\label{eq:KV}
k_{(I)}{}^{a} & = & -\Gamma_{\rm Adj}(u^{-1}(x))^{a}{}_{I}\, ,\\
& & \nonumber \\
\label{eq:Hcompensator}
W_{I}{}^{i} & = & -k_{(I)}{}^{\mu}\vartheta^{i}{}_{\mu} 
-\Gamma_{\rm Adj}(u^{-1}(x))^{i}{}_{I}\, .
\end{eqnarray}


\noindent
\textit{$H$-Covariant Derivatives}

According to the second of Eqs.~(\ref{eq:transformation1}) the
vertical components $\theta^{i}$ transform as an $\mathfrak{h}$-valued
connection. In fact, comparing the Maurer-Cartan equation for the
horizontal components $e^{a}$ with the Cartan structure equation for
the co-frame and (torsionless) spin connection

\begin{equation}
de^{a}-\omega^{a}{}_{b}\wedge e^{b}=0\, ,  
\end{equation}

\noindent 
we find that the spin connection is given by 

\begin{equation}
\label{eq:SC}
\omega^{a}{}_{b} 
=\vartheta^{i}f_{ib}{}^{a}=\vartheta^{i}\Gamma_{\rm Adj}(M_{i})^{a}{}_{b}\, .  
\end{equation}

We use these results to define the {\it $H$-covariant derivative} that
acts on any object that transforms contravariantly $\phi^{\prime} =
\Gamma_{r}(h)\phi$ or covariantly $\psi^{\prime} =
\psi\Gamma_{r}(h^{-1})$ (for instance, $u(x)$ itself) in the
representation $r$ of $H$:

\begin{equation}
\mathcal{D}_{\mu}\phi\equiv \partial_{\mu}\phi 
- \vartheta_{\mu}{}^{i}\Gamma_{r}(M_{i})\phi\, ,
\hspace{1cm}  
\mathcal{D}_{\mu}\psi\equiv \partial_{\mu}\psi 
+\psi\vartheta_{\mu}{}^{i}\Gamma_{r}(M_{i})\, 
\end{equation}

In particular, the Maurer-Cartan equations tell us that 

\begin{equation}
\mathcal{D}_{[\mu}e^{a}{}_{\nu]}=0\, .
\end{equation}

By definition, the Levi-Civit\`a connection is given by 

\begin{equation}
\Gamma_{\mu\nu}{}^{a}\equiv\mathcal{D}_{\mu}e^{a}{}_{\nu}\, .
\end{equation}

Finally, let us introduce the {\it $H$-covariant Lie derivative with
  respect to the Killing vectors $k_{(I)}$}\footnote{$H$-covariant Lie
  derivatives can be defined with respect to any vector, although the
  Lie bracket property Eq.~(\ref{eq:Liebracketproperty}) is only
  satisfied for Killing vectors.  The spinorial Lie derivative
  \cite{kn:Lich,kn:Kos,kn:Kos2} or the Lie-Lorentz derivative that
  naturally appear in calculations of supersymmetry algebras
  \cite{Figueroa-O'Farrill:1999va,Ortin:2002qb} can actually be seen
  as particular examples of this more general operator (see
  e.g.~Ref.~\cite{kn:GoMa}), and, actually, are identical objects when
  acting on Killing spinors of maximally supersymmetric spacetimes, as
  we are going to show.} on objects that transform contravariantly
($\phi$) or covariantly y ($\psi$) in the representation $r$ of $H$:

\begin{equation}
\mathbb{L}_{k_{(I)}}\phi\equiv 
\mathcal{L}_{k_{(I)}}\phi +W_{I}{}^{i}\Gamma_{r}(M_{i})\phi\, ,
\hspace{1cm}  
\mathbb{L}_{k_{(I)}}\psi\equiv 
\mathcal{L}_{k_{(I)}}\psi -\psi W_{I}{}^{i}\Gamma_{r}(M_{i})\, .
\end{equation}

This Lie derivative has, among other properties 

\begin{eqnarray}
\label{eq:Liebracketproperty}
[\mathbb{L}_{k_{(I)}},\mathbb{L}_{k_{(J)}}] & = & 
\mathbb{L}_{[k_{(I)},k_{(J)}]}\, ,  \\
& & \nonumber \\
\mathbb{L}_{k_{(I)}} e^{a} & = & 0\, \\
& & \nonumber \\
\label{eq:lastproperty}
\mathbb{L}_{k_{(I)}} u & = &  
\mathcal{L}_{k_{(I)}}u -u W_{I}{}^{i}M_{i} = T_{I} u\, ,
\end{eqnarray}

\noindent 
where the last property follows from Eqs.~(\ref{eq:Hcompensator}) and
(\ref{eq:esa}).


\section{Killing Spinors in Symmetric Spacetimes}
\label{sec-KSSS}

Most maximally supersymmetric solutions of supergravity theories have
the metric of some symmetric spacetime. In some cases (Minkowski and
$AdS$) the spacetime is also maximally symmetric but in other cases
($AdS\times S$ and $KG$ spacetimes) it is not, but we can always use
the procedure explained in the previous section to construct the
metric, spin connection and Killing vectors. We are going to see,
example by example, that, when we construct in that way the metric,
the Killing spinor equation always takes the form Eq.~(\ref{eq:KSEu}).
It is, nevertheless, convenient to give a brief overview of how we
arrive to the general result. Then, we are going to show how the
general result can be exploited to calculate the commutators of the
supersymmetry algebra.

In all supergravity theories, the Killing spinor equation is of the
the form

\begin{equation}
(\nabla_{\mu} + \Omega_{\mu})\kappa=0\, , 
\end{equation}

\noindent
where the form of $\Omega$ depends on specific details of the theory.
Multiplying by $dx^{\mu}$, it takes the form

\begin{equation}
\left(d -{\textstyle\frac{1}{4}}\omega_{ab}\gamma^{ab} 
+\Omega \right)\kappa=0\, .
\end{equation}

If we construct the symmetric space as in the previous section, then
the spin connection 1-form $\omega_{ab}$ is given by Eq.~(\ref{eq:SC})
and takes values in the vertical Lie subalgebra $\mathfrak{h}$.
Further, 

\begin{equation}
\Gamma_{s}(M_{i})\equiv 
{\textstyle\frac{1}{4}}f_{ia}{}^{b}\gamma_{b}{}^{a}\, ,  
\end{equation}

\noindent
provides a (spinorial) representation of $\mathfrak{h}$ and the
Killing spinor equation becomes

\begin{equation}
\left(d -\vartheta^{i}\Gamma_{s}(M_{i})
+\Omega \right)\kappa=0\, .
\end{equation}

In all the cases that we are going to examine 

\begin{equation}
\Omega = -e^{a}\Gamma_{s}(P_{a})\, ,  
\end{equation}

\noindent
where the matrices $\Gamma_{s}(P_{a})$ are products of a number of
Dirac gamma matrices (and, possibly, of other matrices in extended
supergravities). Thus, on account of the definition of the
Maurer-Cartan 1-forms Eq.~(\ref{eq:M-Cdef}), the Killing spinor equation
can be written in the form Eq.~(\ref{eq:KSEu})

\begin{equation}
\left(d -e^{a}\Gamma_{s}(P_{a}) -\vartheta^{i}\Gamma_{s}(M_{i}) 
\right)\kappa= 
\left(d +\Gamma_{s}(u^{-1}) d \Gamma_{s}(u) \right)\kappa =0\, ,
\end{equation}

\noindent
with

\begin{equation}
\Gamma_{s}(u) = 
e^{x^{1}\Gamma_{s}(P_{1})}\cdots e^{x^{q}\Gamma_{s}(P_{q})}\, ,
\end{equation}

\noindent
and the solution can be written in the form

\begin{equation}
\kappa^{\alpha}=   \Gamma_{s}(u^{-1})^{\alpha}{}_{\beta} \kappa_{0}{}^{\beta}\, ,
\end{equation}

\noindent
for an arbitrary constant spinor $\kappa_{0}{}^{\beta}$ (we have
written explicitly the spinor indices here). Since there will be as
many independent Killing spinors as components has a real
spinor\footnote{We are considering only Majorana spinors.}, we can use
a spinorial index $\alpha$ to label a basis of Killing spinors:

\begin{equation}
\kappa_{(\alpha)}{}^{\beta}=   \Gamma_{s}(u^{-1})^{\beta}{}_{\alpha}\, .
\end{equation}

Killing spinors and Killing vectors are used to find the supersymmetry
algebra of supergravity backgrounds (see,
e.g.~\cite{Gauntlett:1998fz,Figueroa-O'Farrill:1999va,Ortin:2002qb}).
Killing spinors are related to supercharges and Killing vectors to
bosonic charges. The anticommutator of two supercharges gives bosonic
charges and, correspondingly the bilinears
$-i\bar{\kappa}\gamma^{\mu}\kappa$ of Killing spinors are Killing
vectors. To calculate the anticommutator of any two supercharges
$\{Q_{(\alpha)},Q_{(\beta)}\}$ associated to the Killing spinors
$\kappa_{(\alpha)}$ we have to decompose the bilinears into linear
combinations of the Killing vectors $k_{(I)}$

\begin{equation}
-i\bar{\kappa}_{(\alpha)}\gamma^{\mu}\kappa_{(\beta)}\partial_{\mu}
= c_{\alpha\beta}{}^{I}k_{(I)}\, ,  
\end{equation}

\noindent
finding the coefficients $c_{\alpha\beta}{}^{I}$. Now, using the above
general form of the Killing spinors, the bilinears take the form

\begin{equation}
-i\bar{\kappa}_{(\alpha)}\gamma^{\mu}\kappa_{(\beta)}\partial_{\mu} = 
-i\Gamma_{s}(u^{-1})_{\alpha}{}^{\gamma}\mathcal{C}_{\gamma\delta}
(\gamma^{a})^{\delta}{}_{\epsilon} \Gamma_{s}(u^{-1})^{\epsilon}{}_{\beta}\, ,
\end{equation}

\noindent
where $\mathcal{C}$ is the charge conjugation matrix
$\mathcal{C}^{-1}\gamma^{a\, T}\mathcal{C}= -\gamma^{a}$.  Now, in
most cases\footnote{The exception seems to be the Kowalski-Glikman
  H$pp$-wave spacetimes.},the matrices $\gamma^{a}$ happen to be
proportional to the the dual\footnote{It is always possible to find
  the dual of a representation that uses (unitary) gamma matrices.}
$P^{a}$ of a Lie algebra generator $P_{a}$ $\Gamma_{s}(P^{a})$

\begin{equation}
\gamma^{a}=\mathcal{S}\Gamma_{s}(P^{a})\, ,  
\end{equation}

\noindent 
for some matrix $\mathcal{S}$ that depends on the case we are
considering.  The combination
$\tilde{\mathcal{C}}\equiv\mathcal{C}\mathcal{S}$ acts as a charge
conjugation matrix in the subspace spanned by the horizontal
generators in the spinorial representation\footnote{We thank
  P.~Meessen for pointing this out to us.} 

\begin{equation}
\tilde{\mathcal{C}}^{-1}\Gamma_{s}(P^{a})^{T}\tilde{\mathcal{C}}
= -\Gamma_{s}(P^{a})\, ,
\end{equation}

\noindent
so 

\begin{equation}
\Gamma_{s}(u^{-1})^{T}\mathcal{C}\gamma^{a}=   
\Gamma_{s}(u^{-1})^{T}\tilde{\mathcal{C}}\Gamma_{s}(P^{a})=
\tilde{\mathcal{C}}\Gamma_{s}(u)\Gamma_{s}(P^{a})\, .
\end{equation}

\noindent
and, thus, 

\begin{equation}
-i\bar{\kappa}_{(\alpha)}\gamma^{\mu}\kappa_{(\beta)}\partial_{\mu} = 
-i\tilde{\mathcal{C}}_{\alpha\gamma}\Gamma_{s}(u)^{\gamma}{}_{\delta} 
\Gamma_{s}(P^{a})^{\delta}{}_{\epsilon}  
\Gamma_{s}(u^{-1})^{\epsilon}{}_{\beta}e_{a}\, .
\end{equation}

In this expression we can recognize $uP^{a}u^{-1}$ in the spinorial
representation, which is the coadjoint action of the coset element $u$
on $P^{a}$

\begin{equation}
-i\bar{\kappa}_{(\alpha)}\gamma^{\mu}\kappa_{(\beta)}\partial_{\mu} = 
-i\tilde{\mathcal{C}}_{\alpha\gamma}  \Gamma_{s}(T^{I})^{\gamma}{}_{\beta} 
\Gamma_{\rm Adj}(u^{-1})^{a}{}_{I} e_{a} =  
-i\tilde{\mathcal{C}}_{\alpha\gamma}  \Gamma_{s}(T^{I})^{\gamma}{}_{\beta}
k_{(I)}\, , 
\end{equation}

\noindent
where we have used Eq.~(\ref{eq:KV}). The superalgebra structure
constants $c_{\alpha\beta}{}^{I}$ can be readily identified with
$-i\tilde{\mathcal{C}}_{\alpha\gamma}
\Gamma_{s}(T^{I})^{\gamma}{}_{\beta}$.

To complete all the commutation relations of the supersymmetry
algebra, we need the commutators of the bosonic charges and the
supercharges, which are determined by the spinorial or Lie-Lorentz
derivative of the Killing vectors on the Killing spinors
$\mathbb{L}_{k_{(I)}}\kappa_{(\alpha)}$
\cite{Figueroa-O'Farrill:1999va,Ortin:2002qb}, since this operation
preserves the supercovariant derivative (at least in the ungauged
supergravities that we are going to consider) and transforms Killing
spinors into Killing spinors

\begin{equation}
\mathbb{L}_{k_{(I)}}\kappa_{(\alpha)}
=c_{\alpha I}{}^{\beta}\kappa_{(\beta)}\, ,  
\,\,\,\,
\Rightarrow
[Q_{(\alpha)},P_{(I)}]= c_{\alpha I}{}^{\beta} Q_{(\beta)}\, .
\end{equation}

The Lie-Lorentz derivative acting on a (contravariant) spinor $\psi$
is given by \cite{kn:Kos,kn:Kos2}

\begin{equation}
\mathbb{L}_{k_{(I)}}\psi = k_{(I)}{}^{\mu}\nabla_{\mu}\psi 
+{\textstyle\frac{1}{4}} \nabla_{a}k_{(I)}^{b}\gamma^{a}{}_{b}\psi\, .  
\end{equation}

\noindent
On a symmetric space $G/H$, 

\begin{equation}
k_{(I)}{}^{\mu}\nabla_{\mu}\psi = 
k_{(I)}{}^{\mu}\partial_{\mu}\psi -
k_{(I)}{}^{\mu}\vartheta^{i}{}_{\mu}\Gamma_{s}(M_{i})\psi\, ,
\hspace{1cm}
\nabla_{\mu}k_{(I)}^{b} = 
\partial_{\mu}k_{(I)}^{b} -\vartheta^{i}{}_{\mu}f_{ic}{}^{b}k_{(I)}^{c}\, .
\end{equation}

\noindent
Furthermore

\begin{equation}
  \begin{array}{rcl}
\partial_{\mu}k_{(I)}{}^{b} & = & 
-\partial_{\mu}\Gamma_{\rm Adj}(u^{-1})^{b}{}_{I} \\
& & \\
& = & 
\Gamma_{\rm Adj}(u^{-1})^{b}{}_{J}
\partial_{\mu}\Gamma_{\rm Adj}(u)^{J}{}_{K}
\Gamma_{\rm Adj}(u^{-1})^{K}{}_{I}  \\
& & \\
& = & -V^{J}{}_{\mu}f_{JK}{}^{b} \Gamma_{\rm Adj}(u^{-1})^{K}{}_{I}\\
& & \\
& = & -e^{a}{}_{\mu}f_{ai}{}^{b} \Gamma_{\rm Adj}(u^{-1})^{i}{}_{I}
-\vartheta^{i}{}_{\mu}f_{ic}{}^{b} \Gamma_{\rm Adj}(u^{-1})^{c}{}_{I}\\
& & \\
& = & 
e^{a}{}_{\mu}f_{ia}{}^{b} \Gamma_{\rm Adj}(u^{-1})^{i}{}_{I}
+\vartheta^{i}{}_{\mu}f_{ic}{}^{b} k_{(I)}{}^{c}\, ,\\
\end{array}
\end{equation}

\noindent 
so 

\begin{equation}
{\textstyle\frac{1}{4}} \nabla_{a}k_{(I)}^{b}\Gamma^{a}{}_{b}
={\textstyle\frac{1}{4}} f_{ia}{}^{b} \Gamma_{\rm Adj}(u^{-1})^{i}{}_{I}
\gamma^{a}{}_{b}= 
-\Gamma_{\rm Adj}(u^{-1})^{i}{}_{I}\Gamma_{s}(M_{i})\, ,
\end{equation}

\noindent
and

\begin{equation}
  \begin{array}{rcl}
\mathbb{L}_{k_{(I)}}\psi & = &   
k_{(I)}{}^{\mu}\partial_{\mu}\psi -
k_{(I)}{}^{\mu}\vartheta^{i}{}_{\mu}\Gamma_{s}(M_{i})\psi
-\Gamma_{\rm Adj}(u^{-1})^{i}{}_{I}\Gamma_{s}(M_{i}) \psi \\
& & \\
& = &
\mathcal{L}_{k_{(I)}}\psi + W_{I}^{i}\Gamma_{s}(M_{i}) \psi\, .\\
\end{array}
\end{equation}

Then, the Lie-lorentz derivative coincides withe the $H$-covariant
Lie derivative. On the inverse coset representative

\begin{equation}
\mathbb{L}_{k_{(I)}} \Gamma_{s}(u^{-1}) =
-\Gamma_{s}(u^{-1}) 
[\mathbb{L}_{k_{(I)}} \Gamma_{s}(u)] \Gamma_{s}(u^{-1}) = 
-\Gamma_{s}(u^{-1}) \Gamma_{s}(T_{I})\, ,
\end{equation}

\noindent 
on account of Eq.~(\ref{eq:lastproperty}), which implies the commutators

\begin{equation}
\label{eq:QTcommutators}
[Q_{(\alpha)},T_{I}] = -Q_{(\beta)}  \Gamma_{s}(T_{I})^{\beta}{}_{\alpha}\, .
\end{equation}


\subsection{$AdS_{4}$ in $N=1,d=4$ $AdS$ Supergravity}
\label{sec-AdSKS}

$AdS_{4}$ is the maximally supersymmetric vacuum of $N=1,d=4$ $AdS$
supergravity: the integrability conditions of the Killing spinor
equations vanish identically, which implies that 4 independent
solutions exist. They are not hard to find (see
e.g.~Ref~\cite{Lu:1998nu}), but the expressions one gets in most
coordinate systems are difficult to make sense of and they are
difficult to work with to find supersymmetry algebras.

$AdS_{4}$ can be identified with the coset $SO(2,3)/SO(1,3)$. We
introduce $SO(2,3)$ indices $\hat{a},\hat{b},\cdots=-1,0,1,2,3$. The
metric is $\hat{\eta}^{\hat{a}\hat{b}}={\rm diag}(++---)$ and
$\mathfrak{g}=so(2,3)$ the Lie algebra of $SO(2,3)$ can be written in
the general form

\begin{equation}
\label{eq:AdSalgebra}
\left[\hat{M}_{\hat{a}\hat{b}},\hat{M}_{\hat{c}\hat{d}} \right]  =
-\hat{\eta}_{\hat{a}\hat{c}} \hat{M}_{\hat{b}\hat{d}} 
-\hat{\eta}_{\hat{b}\hat{d}} \hat{M}_{\hat{a}\hat{c}}
+\hat{\eta}_{\hat{a}\hat{d}} \hat{M}_{\hat{b}\hat{c}} 
+\hat{\eta}_{\hat{b}\hat{c}} \hat{M}_{\hat{a}\hat{d}}\, .
\end{equation}

We can now perform a $1+4$ splitting of the indices $\hat{a}=(-1,a)\,
,\,\,\, a=0,1,2,3$ and define a new basis 

\begin{equation}
\label{eq:contraction1}
\hat{M}_{ab} = M_{ab}\, ,
\hspace{2cm}
\hat{M}_{a-1} = -g^{-1} P_{a}\, ,
\end{equation}

\noindent 
where we have introduced the dimensionful parameter $g$ related to the
$AdS_{4}$ radius $R$ and to the cosmological constant $\Lambda$ by
\begin{equation}
R=1/g=\sqrt{-3/\Lambda}\, .  
\end{equation}

In terms of the new basis, the $so(2,3)$ algebra reads

\begin{equation}
\begin{array}{rcl}
\left[M_{ab},M_{cd} \right]  & = &
-\eta_{ac}M_{bd} -\eta_{bd} M_{ac}
+\eta_{ad} M_{bc} +\eta_{bc} M_{ad}\, ,\\
& & \\
\left[P_{c},M_{ab}\right]  & = & 
-2P_{[a}\eta_{b]c}\, ,
\hspace{1.5cm}
\left[P_{a},P_{b} \right]  = -g^{2}M_{ab}\, .\\
\end{array}
\end{equation}

The $M_{ab}$s generate the subalgebra $\mathfrak{h}=so(1,3)$ of the
Lorentz subgroup. The complement is $\mathfrak{k}=\{P_{a}\}$ and the
above commutation relations tell us that we have a symmetric pair.
Following the general recipe, we construct the coset representative

\begin{equation}
u(x) = e^{x^{3}P_{3}}  e^{x^{2}P_{2}}e^{x^{1}P_{1}}e^{x^{0}P_{0}}\, ,
\end{equation}

\noindent 
and the Maurer-Cartan 1-forms $e^{a}$, that we are going to use as
Vierbeins are\footnote{In this and similar calculations one has to use
  the formula
  \begin{equation}
  e^{xX}Ye^{-xX}= \cos{x}Y +\sin{x}Z\, ,  
  \end{equation}
where
\begin{equation}
[X,Y]=Z\, ,
\hspace{1cm}  
[Y,Z]=X\, ,
\hspace{1cm}
[Z,X]=Y\, .
\end{equation}
  }


\begin{equation}
e^{0} = -dx^{0}\, ,\,\,\,
e^{1} = -\cos{x^{0}}dx^{1}\, ,\,\,\,
e^{2} = -\cos{x^{0}}{\rm ch}\, x^{1}dx^{2}\, ,\,\,\,
e^{3} = -\cos{x^{0}}{\rm ch}\, x^{1}\, {\rm ch}\, x^{2}dx^{2}\, ,
\end{equation}

\noindent 
and using the Killing metric $(+---)$ we get the $AdS_{4}$ metric 
in somewhat unusual coordinates

\begin{equation}
ds^{2}= (dx^{0})^{2} -\cos^{2}x^{0}\{(dx^{1})^{2} +{\rm ch}^{2}x^{1}
[(dx^{2})^{2} +{\rm ch}^{2}x^{1} (dx^{3})^{2}\}\, .
\end{equation}

We do not need the explicit form of the vertical 1-forms
$\vartheta^{ab}$, but we need to know how they enter the spin
connection. According to the general formula Eq.~(\ref{eq:SC})

\begin{equation}
 \omega^{a}{}_{b} = {\textstyle\frac{1}{4}}e^{cd}f_{cd\, -1b}{}^{-1a}=
 {\textstyle\frac{1}{2}}e\vartheta^{ac}\eta_{cb}\, . 
\end{equation}

The Killing spinor equation is

\begin{equation}
(d -{\textstyle\frac{1}{4}}\omega_{ab}\gamma^{ab} 
-{\textstyle\frac{ig}{2}}\gamma_{a}e^{a}) \kappa=0\, ,  
\end{equation}

\noindent 
and takes immediately the form of Eq.~(\ref{eq:KSEu}) with

\begin{equation}
\Gamma_{s}(P_{a}) = {\textstyle\frac{ig}{2}}\gamma_{a}\, ,
\hspace{1cm}
\Gamma_{s}(M_{ab}) = {\textstyle\frac{1}{2}}\gamma_{ab}\, ,
\end{equation}

\noindent 
and the Killing spinors are of the general form
$\kappa_{(\alpha)}{}^{\beta}=(u^{-1})^{\beta}{}_{(\alpha)}$.

We define the dual generators $\Gamma_{s}(P^{a})$ by

\begin{equation}
  \begin{array}{rcl}
{\rm Tr}\, [\Gamma_{s}(P^{a})\Gamma_{s}(P_{b})]= \delta^{a}{}_{b}\, , 
\,\,\,\,
 & \Rightarrow & 
\Gamma_{s}(P^{a}) = {\textstyle\frac{-i}{2g}}\gamma^{a}\, , \\
& & \\
{\rm Tr}\, [\Gamma_{s}(M^{ab})\Gamma_{s}(P_{cd})]= \delta^{ab}{}_{cd}\, , 
\,\,\,\,
 & \Rightarrow & 
\Gamma_{s}(M^{ab}) = -{\textstyle\frac{1}{2}}\gamma^{ab}\, .\\
\end{array}
\end{equation}

The bilinears are, then ($\mathcal{S}=1$)

\begin{equation}
  \begin{array}{rcl}
-i\bar{\kappa}_{(\alpha)}\gamma^{a}\kappa_{(\beta)} e_{a}
 & =  & 
2g\Gamma_{s}(u^{-1})^{T}\mathcal{C} \Gamma_{s}(P^{a})\Gamma_{s}(u^{-1}) e_{a} 
\\ 
& & \\
& = & g\mathcal{C} \Gamma_{s}(\hat{M}^{\hat{b}\hat{c}}) 
\Gamma_{\rm Adj}(u^{-1})^{a}{}_{\hat{b}\hat{c}}
e_{a} 
\\
& & \\
& = & g\mathcal{C} \Gamma_{s}(\hat{M}^{\hat{b}\hat{c}}) 
k_{(\hat{b}\hat{c})}\, ,\\
\end{array}
\end{equation}

\noindent 
and the anticommutator of the supercharges takes the well-known form

\begin{equation}
\{Q_{(\alpha)},Q_{(\beta)}\} = 
g[\mathcal{C} \Gamma_{s}(\hat{M}^{\hat{a}\hat{b}})]_{\alpha\beta}
\hat{M}_{\hat{a}\hat{b}} = -i(\mathcal{C}\gamma^{a})_{\alpha\beta}P_{a}
-{\textstyle\frac{g}{2}}(\mathcal{C}\gamma^{ab})_{\alpha\beta}M_{ab}\, ,
\end{equation}

\noindent
that reduces to the Poincar\'e supersymmetry algebra in the
$g\rightarrow 0 $ limit.

The commutators $[Q_{(\alpha)},\hat{M}_{\hat{a}\hat{b}}]$ are given by
the general formula (\ref{eq:QTcommutators}):

\begin{equation}
[Q_{(\alpha)},\hat{M}_{\hat{a}\hat{b}}] = 
-Q_{(\beta)}\Gamma_{s}(\hat{M}_{\hat{a}\hat{b}})^{\beta}{}_{\alpha}\, .  
\end{equation}

The generalization to higher dimensions\footnote{Maximally
  supersymmetric $AdS$ vacua arise in gauged supergravities in $d\leq
  7$.} and to spheres, described as cosets $SO(n+1)/SO(n)$ is evident.
As a matter of fact, the coset structure underlies the calculation of
Killing spinors in $S^{n}$ of Ref.~\cite{Lu:1998nu} but only after
this is realized the calculation of bilinears etc.~becomes really
simple.


\subsection{The Robinson-Bertotti Solution in  $N=2,d=4$ Supergravity}
\label{sec-RB}

The Robinson-Bertotti solution of $N=2,d=4$ supergravity
\cite{kn:Rob,kn:Bert} can be obtained as the near-horizon limit of the
extreme Reissner-Nordstr\"om black hole and is known to be maximally
supersymmetric \cite{Gibbons:1984kp,Kallosh:1992gu}, although, to the
best of our knowledge, no explicit expression of its 8 real Killing
spinors is available in the literature. The metric is that of the
direct product of that of $AdS_{2}$ with radius $R_{2}$ and that of
$S^{2}$ with radius $R_{2}$

\begin{equation}
\label{eq:ads2xs2}
\left\{
  \begin{array}{rcl}
ds^{2} & = & R_{2}^{2}\,  d\Pi_{(2)}^{2} 
-R_{2}^{2}\, d\Omega_{(2)}^{2}\, , \\
& & \\
F & = & -\frac{2}{R_{2}}\omega_{AdS_{2}}\, ,\\
  \end{array}
\right.
\end{equation}

\noindent 
where $d\Pi_{(2)}^{2}$ stands for the metric of the $AdS_{2}$
spacetime of unit radius, $d\Omega_{(2)}^{2}$ for the metric of the
unit 2-sphere $S^{2}$ and $\omega_{AdS_{2}}$ for the volume 2-form of
radius $R_{2}$. Both $AdS_{2}$ and $S^{2}$ are symmetric spacetimes
$SO(2,1)/SO(2)$ and $SO(3)/SO(2)$ and we can construct them using the
procedure explained in Section~\ref{sec-symmetric}.

The Lie algebra of $SO(2,1)$ can be written in the form

\begin{equation}
[T_{I},T_{J}]=-\epsilon_{IJK}\mathsf{Q}^{KL}T_{L}\, ,
\,\,\,\,
I,J,\cdots=1,2,3,\, ,  
\hspace{1cm}
\mathsf{Q}={\rm diag}\, (++-)\, , 
\end{equation}

\noindent
and the Lie algebra of $SO(3)$ can be written in the form

\begin{equation}
[\tilde{T}_{I},\tilde{T}_{J}]=-\epsilon_{IJK}\tilde{T}_{K}\, ,
\,\,\,\,
I,J,\cdots=1,2,3,\, .   
\end{equation}

We choose the subalgebra $\mathfrak{h}$ to be generated by $T_{1}$ and
$\tilde{T}_{3}$ so $\mathfrak{k}$ is generated by $T_{2},T_{3}$ and
$,\tilde{T}_{1},\tilde{T}_{2}$. We perform the following redefinitions

\begin{equation}
  \begin{array}{rclrcl}
T_{1} & = & M_{1}\, ,\hspace{1cm} &
\tilde{T}_{1} & = & R_{2}P_{3}\, ,\\
& & & & & \\
T_{2} & = & R_{2}P_{1}\, ,\hspace{1cm} &
\tilde{T}_{2} & = & R_{2}P_{2}\, ,\\
& & & & & \\
T_{3} & = & R_{2}P_{0}\, ,\hspace{1cm} &
\tilde{T}_{3} & = & M_{2}\, ,\\
  \end{array}
\end{equation}

\noindent 
and the coset representative is the product of two mutually commuting
coset representatives $u,\tilde{u}$ with

\begin{equation}
u = e^{R_{2}\phi P_{0}} e^{R_{2}\chi P_{1}}\, ,
\hspace{1cm}  
\tilde{u} = e^{R_{2}\varphi P_{3}} 
e^{R_{2}(\theta-\frac{\pi}{2}) P_{2}}\, .
\end{equation}

\noindent
We get

\begin{equation}
  \begin{array}{rclrcl}
e^{0} & = & -R_{2}{\rm ch}\, \chi\,  d\phi\, ,\hspace{1cm} &  
e^{2} & = & -R_{2}d\theta\, ,\\
& & & & & \\
e^{1} & = & -R_{2}d\chi\, ,&
e^{3} & = & -R_{2}\sin{\theta}d\varphi\, ,\\
& & & & & \\
\vartheta^{1} & = & -{\rm sh}\,\chi\, d\phi\, ,&
\vartheta^{2} & = & -\cos{\theta} d\varphi\, ,\\
  \end{array}
\end{equation}

\noindent 
that lead, using $B={\rm diag}\,(+---)$ to the above $AdS_{2}\times
S^{2}$ metric with

\begin{equation}
\left\{
  \begin{array}{rcl}
d\Pi_{(2)}^{2} & \equiv & {\rm ch}^{2}\chi\, d\phi^{2}-d\chi^{2}\, , \\
& & \\
d\Omega_{(2)}^{2} & \equiv & d\theta^{2}+\sin^{2}{\theta}d\varphi^{2}\, .\\
  \end{array}
\right.
\end{equation}

Contracting with $dx^{\mu}$ the $N=2,d=4$ Killing spinor equation

\begin{equation}
(\nabla_{\mu}
+{\textstyle\frac{1}{8}}\not\!\! F\gamma_{\mu}\sigma^{2})\kappa=0\, ,  
\end{equation}

\noindent 
we immediately see that it takes the form 

\begin{equation}
[d +(u \tilde{u})^{-1}d(u\tilde{u})]=0\, ,
\end{equation}

\noindent 
where the Lie algebra generators are represented by 

\begin{equation}
  \begin{array}{rclrcl}
\Gamma_{s}(P_{0}) & = & \frac{1}{2R_{2}}\gamma^{1}\sigma^{2}\, ,
\hspace{1cm} &  
\Gamma_{s}(P_{2}) & = & 
\frac{1}{2R_{2}}\gamma^{0}\gamma^{1}\gamma^{3}\sigma^{2}\, ,\\
& & & & & \\
\Gamma_{s}(P_{1}) & = & -\frac{1}{2R_{2}}\gamma^{0}\sigma^{2}\, ,&
\Gamma_{s}(P_{3}) & = & 
\frac{1}{2R_{2}}\gamma^{0}\gamma^{1}\gamma^{2}\sigma^{2}\, ,\\
& & & & & \\
\Gamma_{s}(M_{1}) & = & \frac{1}{2}\gamma^{0}\gamma^{1}\, ,&
\Gamma_{s}(M_{2}) & = & \frac{1}{2}\gamma^{2}\gamma^{3}\, .\\
  \end{array}
\end{equation}

The Killing spinors are, then

\begin{equation}
\kappa = (u \tilde{u})^{-1}\kappa_{0}=   
e^{-\frac{1}{2}\phi \gamma^{1}\sigma^{2}}
e^{\frac{1}{2}\chi \gamma^{0}\sigma^{2}}
 e^{-\frac{1}{2}\varphi \gamma^{0}\gamma^{1}\gamma^{2}\sigma^{2}} 
e^{-\frac{1}{2}(\theta-\frac{\pi}{2}) 
\gamma^{0}\gamma^{1}\gamma^{3}\sigma^{2}}\kappa_{0}\, .
\end{equation}

Let us now consider the bilinears $-i\bar{\kappa}\gamma^{\mu}\kappa$
and define the duals $\Gamma_{s}(P^{a})$ by 

\begin{equation}
{\rm Tr}[\Gamma_{s}(P^{a})\Gamma_{s}(P_{b})]  =\delta^{a}{}_{b}\, .
\end{equation}

\noindent 
Then

\begin{equation}
\gamma^{a}=-{\textstyle\frac{4}{R_{2}}} \mathcal{S}\Gamma_{s}(P^{a})\, ,
\hspace{1cm}
\mathcal{S} = \gamma^{0}\gamma^{1}\sigma^{2}\, ,
\end{equation}

\noindent 
and we can see that the modified charge conjugation matrix
$\tilde{\mathcal{C}} =\mathcal{C}\mathcal{S}$ has the required
property

\begin{equation}
\tilde{\mathcal{C}}^{-1}\Gamma_{s}(P_{a})^{T}\tilde{\mathcal{C}}
= -\Gamma_{s}(P_{a})\, ,
\,\,\,\,
\Rightarrow 
(u \tilde{u})^{-1\, T}\tilde{\mathcal{C}} = \tilde{\mathcal{C}}
u \tilde{u}\, ,
\end{equation}

\noindent
that allows us to express the bilinears in the form

\begin{equation}
-i\bar{\kappa}_{(\alpha i)}\gamma^{a}\kappa_{(\beta j)}= 
{\textstyle\frac{4i}{R_{2}}}
\left\{\tilde{\mathcal{C}}[\Gamma_{s}(T^{I})k_{(I)}
+\Gamma_{s}(\tilde{T}^{I})\tilde{k}_{(I)}]
 \right\}_{(\alpha i\, \beta j)}\, ,  
\end{equation}

\noindent
where the $k_{(I)}$s are the Killing vectors of $AdS_{2}$ and the
$\tilde{k}_{(I)}$s are those of $S^{2}$. This translates into the
anticommutator

\begin{equation}
\{Q_{(\alpha i)},Q_{(\beta j)}\} =
-i\delta_{ij}(\mathcal{C}\gamma^{a})_{\alpha\beta}P_{a}
+{\textstyle\frac{i}{R_{2}}}\mathcal{C}_{\alpha\beta}  \epsilon_{ij} M_{1}
+{\textstyle\frac{1}{R_{2}}}(\mathcal{C}\gamma_{5})_{\alpha\beta}  
\epsilon_{ij} M_{2}\, .
\end{equation}

The commutators of the supercharges and the bosonic generators are
given by the general formula (\ref{eq:QTcommutators}).


\subsection{Other $AdS\times S$ Solutions}
\label{sec-AdSxS}

There are some other maximally supersymmetric vacua of supergravity
theories with metrics which are the direct product of $AdS_{n}$ and
$S^{m}$ spacetimes. They typically arise in the near-horizon limit of
$p$-brane solutions that preserve only a half of the supersymmetries
\cite{Gibbons:sv} and can be used in Freund-Rubin compactifications
\cite{Freund:1980xh}, with $S^{m}$ as internal space, to get gauged
supergravities in $n$ dimensions with gauge group $SO(n+1)$. The known
cases are $AdS_{4}\times S^{7}$ and $AdS_{7}\times S^{4}$in $N=1,d=11$
supergravity, $AdS_{5}\times S^{5}$ in $N=2B,d=10$ supergravity,
$AdS_{3}\times S^{3}$ in $N=2,d=6$ supergravity, $AdS_{2}\times S^{3}$
\cite{Chamseddine:1996pi} and $AdS_{3}\times S^{2}$
\cite{Gibbons:1994vm} in $N=2,d=5$ supergravity and the
Robinson-Bertotti solution $AdS_{2}\times S^{2}$ in $N=2,d=4$ that we
have just studied and that can be taken as prototype.

The Killing spinors of all these solutions can be obtained in similar
forms. The only complications that arise are due to the
symplectic-Majorana nature of supergravity spinors in $4<d<8$.  We are
going to see next how the Killing spinors and vectors the
supersymmetry algebras of $AdS_{4}\times S^{7}$ and $AdS_{7}\times
S^{4}$in $N=1,d=11$ supergravity and $AdS_{5}\times S^{5}$ in
$N=2B,d=10$ supergravity can be quickly obtained.


\subsubsection{$AdS_{4}\times S^{7}$ in $N=1,d=11$ Supergravity}

This solution is given by

\begin{equation}
\label{eq:ads4xs7}
\left\{
  \begin{array}{rcl}
ds^{2} & = & R_{4}^{2}\,  d\Pi_{(4)}^{2} 
-(2R_{4})^{2}\, d\Omega_{(7)}^{2}\, , \\
& & \\
G & = & \frac{3}{R_{4}}\omega_{AdS_{4}}\, ,
\,\,\,\,
\Rightarrow 
G_{0123}=\frac{3}{R_{4}}\, ,\\
  \end{array}
\right.
\end{equation}

\noindent 
where $d\Pi_{(4)}^{2}$ stands for the metric of the $AdS_{4}$
spacetime of unit radius, $d\Omega_{(7)}^{2}$ for the metric of the
unit 7-sphere $S^{7}$ and $\omega_{AdS_{4}}$ for the volume 4-form of
radius $R_{4}$.

We construct $AdS_{4}$ as in Section~\ref{sec-AdSKS} with $g=1/R_{4}$
and this gives us the first four Elfbeins $e^{a}$ associated to the
generators $P_{a}$ $a=0,1,2,3$ and the first 6 1-forms
$\vartheta^{ab}= -\vartheta^{ba}$ associated to the first 4 generators
of the 11-dimensional Lorentz group $M_{ab}$ $a,b=0,1,2,3$. The
detailed expressions of these 1-forms is really not necessary.

To construct the sphere of radius $2R_{4}$ we split the $SO(8)$ Lie
algebra generators

\begin{equation}
\label{eq:Salgebra}
\left[\tilde{M}_{\tilde{a}\tilde{b}},\tilde{M}_{\tilde{c}\tilde{d}} \right]  =
\delta_{\tilde{a}\tilde{c}} \tilde{M}_{\tilde{b}\tilde{d}} 
+\delta_{\tilde{b}\tilde{d}} \tilde{M}_{\tilde{a}\tilde{c}}
-\delta_{\tilde{a}\tilde{d}} \tilde{M}_{\tilde{b}\tilde{c}} 
-\delta_{\tilde{b}\tilde{c}} \tilde{M}_{\tilde{a}\tilde{d}}\, .
\end{equation}

\noindent
into 

\begin{equation}
\tilde{M}_{8i}= 2R_{4}P_{i}\, ,
\hspace{1cm} 
\tilde{M}_{ij}=M_{i+3\, j+3}\, ,
\,\,\,\, i,j=1,\cdots 7\, , 
\end{equation}

\noindent 
and provide the last 7 $P_{a}$'s and Lorentz generators $M_{ab}$
$a,b=4,\dots,8$. The standard procedure also gives us the associated 7
Elfbeins $e^{a}$ and 1-forms $\vartheta^{ab}$ $a,b=4,\dots,8$. Again,
the detailed expressions are not necessary. The metric in
Eq.~(\ref{eq:ads4xs7}) is obtained using the Killing metric of both
factors $(+-\cdots -)$.

The general arguments given at the beginning of this section ensure that

\begin{equation}
dx^{\mu}\nabla_{\mu} = d - \sum_{a<b}\vartheta^{ab}\Gamma_{s}(M_{ab})\, ,  
\hspace{1cm}
\Gamma_{s}(M_{ab}) ={\textstyle\frac{1}{2}}\Gamma_{ab}\, ,
\end{equation}

\noindent
and a straightforward calculation gives for the second piece
of the Killing spinor equation

\begin{equation}
{\textstyle\frac{i}{288}}\left(\Gamma^{abcd}{}_{f} e^{f}
-8\Gamma^{abc}e^{d} \right)G_{abcd}=
-e^{a}\Gamma_{s}(P_{a})\, ,
\end{equation}

\noindent 
where 

\begin{equation}
\Gamma_{s}(P_{a}) =
\left\{
  \begin{array}{ccc}
\frac{i}{2R_{4}}\Gamma^{0123}\Gamma_{a}\, ,& \hspace{1cm} & a\leq 3\, ,\\
& & \\
-\frac{i}{4R_{4}}\Gamma^{0123}\Gamma_{a}\, ,& \hspace{1cm} & a> 3\, .\\
  \end{array}
\right.  
\end{equation}

\noindent
The Killing spinor equation takes the general form Eq.~(\ref{eq:KSEu})
and is solved as usual. The specific form of the solution depends on
the specific choice of coset representative, but it is unimportant in
what follows.

Now, let us consider the bilinears
$-i\bar{\kappa}_{(\alpha)}\Gamma^{a}\kappa_{(\beta)}$. Let us define
generators $\Gamma_{s}(P^{a})$ dual to the $\Gamma_{s}(P_{a})$ that
are exponentiated to construct the coset representative 

\begin{equation}
{\rm Tr}\, [ \Gamma_{s}(P^{a}) \Gamma_{s}(P_{b})] = \delta^{a}{}_{b}\, .
\end{equation}

They are given by 

\begin{equation}
  \begin{array}{rcl}
\Gamma_{s}(P^{a}) & = &
\left\{
  \begin{array}{ccc}
-\frac{iR_{4}}{16}\Gamma^{0123}\Gamma^{a}\, ,& \hspace{1cm} & a\leq 3\, ,\\
& & \\
-\frac{iR_{4}}{8}\Gamma^{0123}\Gamma^{a}\, ,& \hspace{1cm} & a> 3\, .\\
  \end{array}
\right.\, ,\\
& & \\
\Gamma_{s}(M^{ab}) & = & -\frac{1}{16}\Gamma^{ab}\, .\\
\end{array}
\end{equation}

The gamma matrices that appear in the bilinears are related to these by

\begin{equation}
  \begin{array}{rcl}
\Gamma^{a} & = & {\textstyle\frac{-16i}{R_{4}}}
\mathcal{S}\Gamma_{s}(P^{a})\, ,\hspace{1cm}a\leq 3\, ,\\
& & \\
\Gamma^{a} & = & {\textstyle\frac{-8i}{R_{4}}}
\mathcal{S}\Gamma_{s}(P^{a})\, ,\hspace{1cm}a> 3\, ,\\
  \end{array}
\hspace{1cm}
\mathcal{S}=\Gamma^{0123}\, ,
\end{equation}

\noindent
and, since the modified charge conjugation matrix $\tilde{\mathcal{C}}
=\mathcal{C}\mathcal{S}$ has the required property

\begin{equation}
\label{eq:property}
\tilde{\mathcal{C}}^{-1}\Gamma_{s}(P^{a})^{T}\tilde{\mathcal{C}} 
= -\Gamma_{s}(P^{a})\, ,
\end{equation}

\noindent
the bilinears can be written in the form (suppressing the indices
$\alpha,\beta$)

\begin{equation}
-i\bar{\kappa}\Gamma^{a}\kappa = {\textstyle\frac{-8}{R_{4}}} 
\tilde{\mathcal{C}}[\Gamma_{s}(\hat{M}^{\hat{a}\hat{b}})k_{(\hat{a}\hat{b})}
+{\textstyle\frac{1}{2}}\Gamma_{s}(\tilde{M}^{\tilde{a}\tilde{b}})
k_{(\tilde{a}\tilde{b})}]\, ,
\end{equation}

\noindent
where hatted generators and Killing vectors belong to the $AdS_{4}$ factor
and the tilded ones to the $S^{7}$ factor. The
anticommutator of two supercharges can be immediately read in this
expression and the commutator of supercharges and bosonic charges is
given by the general formula Eq.~(\ref{eq:QTcommutators}).


\subsubsection{$AdS_{7}\times S^{4}$ in $N=1,d=11$ Supergravity}

This solution is given by

\begin{equation}
\label{eq:ads7xs4}
\left\{
  \begin{array}{rcl}
ds^{2} & = & R_{7}^{2}\,  d\Pi_{(7)}^{2} 
-(R_{7}/2)^{2}\, d\Omega_{(4)}^{2}\, , \\
& & \\
G & = & \frac{6}{R_{7}}\omega_{S^{4}}\, ,
\,\,\,\,
\Rightarrow 
G_{789\, 10}=\frac{6}{R_{7}}\, ,\\
  \end{array}
\right.
\end{equation}

\noindent 
where we use the same notation as in the preceding cases and
$\omega_{S^{4}}$ stands for the volume of the sphere of radius
$R_{7}/2$. The definitions of the $P_{a}$ and $M_{ab}$ generators and
the construction of the Elfbeins etc.~is almost identical to that of
the preceding case and we immediately arrive at

\begin{equation}
dx^{\mu}\nabla_{\mu} = d - \sum_{a<b}\vartheta^{ab}\Gamma_{s}(M_{ab})\, ,  
\hspace{1cm}
\Gamma_{s}(M_{ab}) ={\textstyle\frac{1}{2}}\Gamma_{ab}\, .
\end{equation}

The 1-forms $\vartheta^{ab}$ have a different form now, but we do not
need to know it.  The second piece of the Killing spinor equation
takes the form

\begin{equation}
{\textstyle\frac{i}{288}}\left(\Gamma^{abcd}{}_{f} e^{f}
-8\Gamma^{abc}e^{d} \right)G_{abcd}=
-e^{a}\Gamma_{s}(P_{a})\, ,
\end{equation}

\noindent 
where now 

\begin{equation}
\Gamma_{s}(P_{a}) =
\left\{
  \begin{array}{ccc}
\frac{i}{2R_{7}}\Gamma^{789\, 10}\Gamma_{a}\, ,& \hspace{1cm} & a\leq 6\, ,\\
& & \\
-\frac{i}{R_{4}}\Gamma^{789\, 10}\Gamma_{a}\, ,& \hspace{1cm} & a> 6\, .\\
  \end{array}
\right.  
\end{equation}

The Elfbeins are also different, but, yet again, we do not need to
know their detailed expressions. The dual generators are defined as usual
and are given by

\begin{equation}
  \begin{array}{rcl}
\Gamma_{s}(P^{a}) & = &
\left\{
  \begin{array}{ccc}
-\frac{iR_{7}}{16}\Gamma^{789\, 10}\Gamma^{a}\, ,& \hspace{1cm} & a\leq 6\, ,\\
& & \\
-\frac{iR_{7}}{32}\Gamma^{789\, 10}\Gamma^{a}\, ,& \hspace{1cm} & a> 6\, .\\
  \end{array}
\right.\, ,\\
& & \\
\Gamma_{s}(M^{ab}) & = & -\frac{1}{16}\Gamma^{ab}\, .\\
\end{array}
\end{equation}

\noindent
and

\begin{equation}
  \begin{array}{rcl}
\Gamma^{a} & = & {\textstyle\frac{16i}{R_{7}}}
\mathcal{S}\Gamma_{s}(P^{a})\, ,\hspace{1cm}a\leq 6\, ,\\
& & \\
\Gamma^{a} & = & {\textstyle\frac{-32i}{R_{7}}}
\mathcal{S}\Gamma_{s}(P^{a})\, ,\hspace{1cm}a> 6\, ,\\
  \end{array}
\hspace{1cm}
\mathcal{S}=\Gamma^{789\, 10}\, .
\end{equation}

The modified charge conjugation matrix has the property
Eq.~(\ref{eq:property}) and we get, suppressing again $\alpha\beta$
indices

\begin{equation}
-i\bar{\kappa}\Gamma^{a}\kappa = {\textstyle\frac{-8}{R_{7}}} 
\tilde{\mathcal{C}}[\Gamma_{s}(\hat{M}^{\hat{a}\hat{b}})k_{(\hat{a}\hat{b})}
-2\Gamma_{s}(\tilde{M}^{\tilde{a}\tilde{b}})
k_{(\tilde{a}\tilde{b})}]\, ,
\end{equation}

\noindent
where hatted generators and Killing vectors belong to the $AdS_{7}$
factor and the tilded ones to the $S^{4}$ factor. Again, the
anticommutator of two supercharges can be immediately read in this
expression and the commutator of supercharges and bosonic charges is
given by the general formula Eq.~(\ref{eq:QTcommutators}).


\subsubsection{$AdS_{5}\times S^{5}$ in $N=2B,d=10$ Supergravity}

The solution is given in the string frame by

\begin{equation}
  \left\{
  \begin{array}{rcl}
  ds^2&=&R_5^2d\Pi_{(5)}^2-R_5^2d\Omega_{(5)}^2\, , \\ \\ 
  G^{(5)}&=&\frac{4e^{-\varphi_0}}{R_5}(\omega_{AdS_5}+\omega_{S^5})\, , 
  \ \ \ \Rightarrow 
  G^{(5)}_{01234}=G^{(5)}_{56789}=\frac{4e^{-\varphi_0}}{R_5}\, , \\ \\ 
  \varphi&=&\varphi_0\, .  
  \end{array}
  \right.
\end{equation}
This case is exactly analogous to the previous ones. The  
normalization in the splitting of the generators of $SO(2,4)$ and 
$SO(6)$ is now, respectively:
\begin{equation}
  \begin{array}{rcl}
  \hat{M}_{a-1}&=&-R_5P_a\, , \ \ \ (a=0,...,4) \\ \\ 
  \tilde{M}_{6a}&=&R_5P_a\, , \ \ \ (a=5,...,9)\, .
  \end{array}
\end{equation}

Once again we do not need to know the explicit form of the Zehnbeins. 
From the covariant derivative term in the gravitino supersymmetry 
transformation (the variation of the dilatino vanishes automatically) 
we get the generators of $SO(1,4)$ and $SO(5)$ in the spinor 
representation. From the remaining piece: 

\begin{equation}
  -\textstyle\frac{1}{16\cdot 5!}e^{\varphi_0}G^{(5)}_{bcdef}
  \Gamma^{bcdef}\Gamma_ai\sigma^2 = -e^a\Gamma_{s}(P_a)\, ,
\end{equation}

\noindent
we read the spinor representation for the generators
$P_a$\footnote{Here there is another (completely equivalent, since we
  are dealing with chiral spinors) possibility, consisting in replacing
  $\Gamma^{01234}$ by $-\Gamma^{56789}$. }

\begin{equation}
  \Gamma_{s}(P_{a})=\left\{
  \begin{array}{c}
  \frac{i}{2R_5}\sigma^{2}\Gamma^{01234}\Gamma_a\, , 
  \ \ \ (a=0,\cdots,4) \\ \\ 
  -\frac{i}{2R_5}\sigma^{2} \Gamma^{01234}\Gamma_a\, .
  \ \ \ (a=5,\cdots,9)
  \end{array} 
  \right. 
\end{equation}

The dual generators are 

\begin{equation}
\Gamma_{s}(P^{a}) = {\textstyle\frac{iR_{5}}{32}} \sigma^{2}
\Gamma^{01234}\Gamma^{a}\, ,
\,\,\,\,
\Rightarrow
\Gamma^{a} = {\textstyle\frac{32i}{R_{5}}} \mathcal{S}\Gamma_{s}(P^{a})\, ,
\hspace{1cm}
\mathcal{S}=\sigma^{2}\Gamma^{01234}\, , 
\end{equation}

\noindent
and the modified charge conjugation matrix has the required property
Eq.~(\ref{eq:property}) that leads to

\begin{equation}
-i\bar{\kappa}\Gamma^{a}\kappa = {\textstyle\frac{32}{R_{5}}} 
\tilde{\mathcal{C}}[\Gamma_{s}(\hat{M}^{\hat{a}\hat{b}})k_{(\hat{a}\hat{b})}
+\Gamma_{s}(\tilde{M}^{\tilde{a}\tilde{b}})
k_{(\tilde{a}\tilde{b})}]\, .
\end{equation}


\subsection{H$pp$-wave Spacetimes and the $KG4,5,6,10,11$ Solutions}
\label{sec-HppKG}

Although maximally supersymmetric $pp$-wave solutions were discovered
long time ago by Kowalski-Glikman in $N=2,d=4$ and $N=1,d=11$
supergravity \cite{Kowalski-Glikman:wv,Kowalski-Glikman:1985im}, only
recently they have received wide attention. This renewed interest has
been accompanied with the discovery of new maximally supersymmetric
solutions of the same kind (henceforth $KG$ solutions) in $N=2B,d=10$
supergravity \cite{Blau:2001ne} and in $N=2,d=5,6$ supergravities
\cite{Meessen:2001vx}, and by the realization that they can be
obtained by taking a Penrose limit \cite{kn:Pen6,Gueven:2000ru} of the
known $AdS\times S$ maximally supersymmetric solutions
\cite{Blau:2002dy,Blau:2002rg}.

The $KG$ solutions are particular examples of homogeneous $pp$-wave
spacetimes (H$pp$-waves), symmetric spacetimes to which we can apply
our formalism. Let us review briefly the coset construction that leads
to them \cite{kn:CaWa,Figueroa-O'Farrill:2001nz}.

The generators of $\mathfrak{g}$ in H$pp$-wave spacetimes are
$\{T_{-},T_{+},T_{i},T_{\star i}\}$ $i=1,\cdots,d-2$ and their
non-vanishing Lie brackets are

\begin{equation}
[T_{-},T_{i}]= T_{\star i}\, ,
\hspace{.7cm}  
[T_{-},T_{\star i}]= A_{ij}T_{j}\, ,
\hspace{.7cm}  
[T_{i},T_{\star j}] = A_{ij}T_{+}\, , 
\hspace{.7cm}  
A_{ij}=A_{ji}\, .
\end{equation}

$T_{+}$ is central in this Lie algebra.  The subalgebra $\mathfrak{h}$
is generated by the $T_{\star i}\equiv M_{i}$ and $\mathfrak{k}$ is
generated by $T_{-}\equiv P_{-}\, ,\,\,\, T_{+}\equiv P_{+}\, ,\,\,\,
T_{i}\equiv P_{i}$, and the coset representative is chosen to be

\begin{equation}
u = e^{x^{-}P_{-}}e^{x^{+}P_{+}}e^{x^{i}P_{i}}\, ,  
\end{equation}

\noindent
which lead to the Maurer-Cartan 1-form

\begin{equation}
V = u^{-1}du = -dx^{-}P_{-} 
-(dx^{+} +{\textstyle{\frac{1}{2}}}x^{i}x^{j}A_{ij}dx^{-})P_{+} 
-dx^{i}P_{i}
-x^{i}dx^{-}M_{i}\, .
\end{equation}

Since $\mathfrak{g}$ is not semisimple, its Killing metric is singular
and cannot be used to construct a $G$-invariant metric. Instead, we
choose\footnote{The metric $B_{+-}=1\, ,\,\,B_{ij}= -\delta_{ij}$ is
  not invariant under the action of $\mathfrak{h}$ on $\mathfrak{k}$.
  Thus, we are forced to work with mostly plus signature in this
  section.}

\begin{equation}
B_{+-}=1\, ,
\hspace{1cm}
B_{ij}= +\delta_{ij}\, ,
\end{equation}

\noindent
and we get the general H$pp$-wave metric

\begin{equation}
ds^{2} = 2dx^{-}(dx^{+} +{\textstyle{\frac{1}{2}}}x^{i}x^{j}A_{ij}dx^{-})
+dx^{i}dx^{i}\, .  
\end{equation}

Different H$pp$-wave metrics are characterized by the matrix $A_{ij}$
up to $SO(d-2)$ rotations. On the other hand (and this is an important
difference with the previous cases), the H$pp$-wave metric can have
more isometries: all possible rotations of the $x^{i}$ that preserve
the matrix $A_{ij}$. These rotations do not belong to $\mathfrak{g}$
and the corresponding Killing vectors cannot be found by applying
Eq.~(\ref{eq:KV}). 

Let us now consider the $KG11$ solution. Its metric is of the above
general H$pp$-wave form, with $A_{ij}$ and the 4-form field strength
given by

\begin{equation}
G_{-123}= \lambda\, ,
\hspace{1cm}
A_{ij}= 
\left\{
  \begin{array}{ccl}
-\frac{1}{9}\lambda^{2}\delta_{ij} & \hspace{.6cm} & i,j=1,2,3\, ,\\
& & \\
-\frac{1}{36}\lambda^{2}\delta_{ij} & \hspace{.6cm} & i,j=4,\cdots,9\, .\\
  \end{array}
\right.
\end{equation}

This solution is additionally invariant under rotations in the
subspaces parametrized by $x^{1},x^{2},x^{3}$ and
$x^{4},\cdots,x^{9}$. Let us now consider the Killing spinor equation.
According to the general construction, we only need to compute the
$\Omega$ part that involves the 4-form field strength. This can be
written in the form $-e^{a}\Gamma_{s}(P_{a})$ with 

\begin{equation}
  \begin{array}{rcl}
\Gamma_{s}(P_{-}) & = & 
\frac{\lambda}{12}(\Gamma^{-}\Gamma^{+}+1)\Gamma^{123}\, , \\
& & \\
\Gamma_{s}(P_{+}) & = & 0\, ,\\
& & \\  
\Gamma_{s}(P_{i}) & = &
\left\{
  \begin{array}{c}
-\frac{\lambda}{6}\Gamma^{-123}\Gamma_{i}\, ,\,\,\,\, i=1,2,3\, ,\\
\\
-\frac{\lambda}{12}\Gamma^{-123}\Gamma_{i}\, ,\,\,\,\, i=4,\cdots,9\, ,\\
    \end{array}
\right. 
\end{array}
\end{equation}

\noindent
and the Killing spinor has the same form as usual, the only difference
being that one of the $P_{a}$ generators ($P_{+}$)is represented by
zero and does not contribute to the coset representative $u$. The
general formula Eq.~(\ref{eq:QTcommutators}) can be used to calculate
the commutators of supercharges and bosonic generators. We see that
$P_{+}$ is a central charge also in the superalgebra
\cite{Figueroa-O'Farrill:2001nz}. The calculation of the
anticommutators of supercharges is more complicated, though, basically
because we can construct duals 

\begin{equation}
\Gamma_{s}(P^{a})\sim \Gamma^{+123}\Gamma^{a}\, ,  
\end{equation}

\noindent
but the matrix $\Gamma^{+123}$ is singular and the relation cannot be
inverted. This is related to the existence of the extra rotational
Killing vectors $k_{(ij)}$ that do appear in the bilinear
$-i\bar{\kappa}_{(\alpha)}\Gamma^{a}\kappa e_{a}$
\cite{Figueroa-O'Farrill:2001nz}. The above equation can in fact be
used to relate the Killing vectors $k_{(I)}$ to some of all the
possible bilinears. The additional Killing vectors $k_{(ij)}$ appear
in the other bilinears (associated to the anticommutators
$\{Q_{-},Q_{-}\}$ in the notation of
Ref.~\cite{Figueroa-O'Farrill:2001nz}).


\section{Conclusions}
\label{sec-conclusions}

In this paper we have checked in almost all known maximally
supersymmetric backgrounds that the Killing spinor equation can be set
in the form Eq.~(\ref{eq:KSEu}) and we have shown how this can be
exploited to calculate their supersymmetry algebras using results from
the theory of symmetric spaces. 

There are two exceptional cases: the $KG$ spaces, for which it is not
easy to compute all the possible anticommutators
$\{Q_{(\alpha)},Q_{(\beta)}\}$ and the maximally supersymmetric
solution that can be obtained by taking the near-horizon limit of the
5-dimensional extreme rotating black hole
\cite{Kallosh:1996vy,Gauntlett:1998fz,Cvetic:1998xh} whose description
as symmetric space is not known.

The obvious extension of this work is to backgrounds with less
supersymmetry, like those that can be obtained by replacing the sphere
in $AdS\times S$ solutions by another homogeneous space with the right
curvature \cite{Castellani:1983yg,Duff:1995wk,Castellani:1998nz}. Work
in this direction is in progress.

{\bf Note added in Proof:} the authors are indebted to the Referee for
pointing out to us the actual group-theoretical reason due to which,
in the case of H{\em pp}-wave geometries, not all the Killing vectors
can be found by using eq.(1.20). This is because the complete isometry
group of the considered H{\em pp}-wave solutions is a semi-direct
product of the group defined in eq.(2.81) and the group of additional
isometries and, since the coset space only belongs to one of the
factors in this semi-direct product, (1.20) does not produce the
Killing vectors corresponding to the other factor (the rotational
isometries of the wave-front). \\

\vspace{1cm}

\section*{Acknowledgments}

The authors would like to thank P.~Meessen for many useful
discussions.  T.O.~would like to thank M.M.~Fern\'andez for her
continuous support.  This work has been partially supported by the
Spanish grant FPA2000-1584.



\end{document}